# DNA Encoded Elliptic Curve Cryptography System for IoT Security

***Prokash Barman*[a*]*, Banani Saha*[b]**

[a]*Department of Computer Science & Engg, University of Calcutta, Technology Campus, JD-2, Sector 3, Salt Lake City, Kolkata 700 098, India*
[b]*Department of Computer Science & Engg, University of Calcutta, Technology Campus, JD-2, Sector 3, Salt Lake City, Kolkata 700 098, India*

Abstract: In the field of Computer Science and Information Technology Internet of Things (IoT) is one of the emerging technologies. In IoT environment several devices are interconnected and transmit data among them. There may be some security vulnerability arise within the IoT environment. Till date, IoT has not been widely accepted due to its security flaws. Hence to keep the IoT environment most robust, we propose a stable security framework of IoT with Elliptic Curve Cryptography (ECC) using DNA Encoding. The ECC is most lightweight cryptography technique among other well known public key cryptography techniques. To increase encryption complexity, DNA encoding mechanism of DNA computing with ECC is preceded.

Keywords: IoT, ECC, DNA, Encoding, Cryptography

## 1. Introduction

A collection of interconnected people, services, objects and devices which can communicate, share data and information to achieve common goals in different areas and applications is termed as Internet of things (IoT). European Research Cluster on the Internet of Things (IERC) states that "Internet of Things is a dynamic global network infrastructure with self-configuring capabilities based on standard and interoperable communication protocols where physical and virtual things have identities, physical attributes and virtual personalities and use intelligent interface and are seamlessly integrated into the information network" [O. Vermesan and P. Friess, 2013]. In a nutshell, IoT is characterized by the real world of smart objects with limited storage and processing power [R. Roman, P. Najera & J. Lopez,2011]. IoT may be implemented in different domains like as agriculture, transportation, healthcare, energy production and distribution, and many other areas which require internet types of things to communicate over the Internet to perform tasks intelligently without human intervention. An Identity Management (IM) approach to be followed in a collection of similar and heterogeneous devices which are participating in IoT operations. In an IoT region, each entity must have a unique identifier and the IoT region may be defined by an IP address.

The utilization of IoT may transform the way of our living behavior by involving intelligent devices around us to perform daily tasks and activities with minimal human intervention. The terms which are used in relevance to IoT are Smart homes, smart cities, smart transportation and infrastructure etc. There are various application domains for IoT, ranging from personal to enterprise environments [M. Abomhara and G. M. Koien, 2014]. IoT is also applied in the transportation sector, in which various smart cars, smart roads, and smart traffic signals serve the purpose of safe and suitable transportation facilities. The enterprise and industries domain of IoT include the applications used in finance, banking, marketing etc. to enable different activities in organizations. The last application domain is the service and utility monitoring sector which includes agriculture, breeding, energy management, and recycling operations, etc. [M. Abomhara and G. M. Koien, 2014].

### *1.1. Need of IoT Security*

The expansion of internet-connected data, devices, applications and users has exploded exponentially. IoT is carrying over into such a wide array of products and services: mobile devices, wearable, medical devices; everything under the sun can now be connected to the internet. With so many organizations offering internet-connected products and services, an equal effort should be made to prevent unauthorized access or data breaches via IoT devices. Simply put, with an IoT device comes the risk of malicious actors taking command of internet-connected controls. Given the risk of compromising customer data or access to private systems, the effort to secure IoT devices is as important as making IoT devices available for use. It's a well-known fact within the information security industry that hackers are getting more sophisticated every day, finding new and unique ways to exploit security vulnerabilities. Furthermore, in some cases, it's becoming less expensive to exploit security vulnerabilities. As the hackers are becoming more skilled, corporations are also increasing IoT-related activity, providing more opportunities for exploitation. There are countless internet-facing devices on the market today that connect and share sensitive data, but have rudimentary security controls, such as a solitary password authentication. Given the risky environment of IoT today, a single password is no longer sufficient for authentication. Multiple layers of identity authentication across a user, device application and data need to be in place to adequately mitigate risk and prevent breaches.

Fundamentally, the main contributors to the importance of IoT security can be condensed into two key considerations:

**"The growth of internet-connected devices, more devices=more vulnerabilities"**
**"Not all devices and applications are built with security in mind, which increases the number of attack vectors and potential targets for hackers. How to get it"** [Blog: why-is-iot-security-so-critical]

IoT has not been implemented in all the fields due to its inadequate security architecture. The review of literature articulates the state of the art in this field with its diversified application domains in creating the smart environment and the challenges in deploying this new scenario to all the fields. The significant challenges to be resolved with the existing smart applications are inter-operability, security, QoS, load balancing, mobility, IPv6 deployment, data management solution and acceptability of IoT applications by users and citizens. Several research works have been carried out to address these issues faced in creating the smart environment and to extend the same for all application domains and so far some proposed work seems to be integrated and secured. In the paper [Daisy Premila Bai T, Albert Rabara S & Vimal Jerald M] a novel Elliptic Curve Cryptography based security framework for Internet of Things and cloud computing has been proposed. In this paper we proposed a multi level security mechanism with DNA encoded Elliptic Curve Cryptography to secure IoT. This paper is organized in follows the following manner. Section 2 briefs the Security Challenges within IoT Systems. Section 3 describes solution for IoT Security implementation. Section 4 presents proposed security mechanism i.e. DNA encoded ECC for IoT security. Section 5 presents the need of resilient mechanism in IoT. In Section 6 we conclude the paper.

Elliptic curve cryptography is a public key cryptosystem developed by Neil Kobiltz and Victor Miller in 19th century [N. Koblitz, 1987][V.S. Miller, 1985]. The security strength of ECC depends on the difficulty of Elliptic Curve Discrete Logarithm Problem (ECDLP) [Moncef Amara and Amar Siad, 2011]. ECC adopts scalar multiplication, which includes point doubling and adding operation which is computationally more efficient than RSA public key cryptography exponentiation. The complexity of ECC puts the attacker in difficulty to understand the ECC and to break the security key. The security level achieved by RSA using 1024 bit key can be obtained with 160 bit key by ECC. So ECC is well suited for resource constraint devices like smart cards, mobile devices, etc. [Sandeep S. Kumar, 2006].

## 2. Security Challenges within IoT Systems

In many cases, a major commotion of the conventional model brings it's own set of challenges. In designing and building of IoT devices or systems, some security challenges and considerations are as the following:

1. Typically small, inexpensive devices with little to no physical security
2. Computing platforms, constrained in memory and compute resources, may not support complex and evolving security algorithms due to the following factors:
   a) Limited security compute capabilities
   b) Encryption algorithms need higher processing power
   c) Low CPU cycles vs. effective encryption
3. Designed to operate autonomously in the field with no backup connectivity if primary connection is lost.
4. Mostly installed prior to network availability which increases the overall on-boarding time.
5. Requires secure remote management during and after on boarding.
6. Scalability and management of billions of entities in the IoT ecosystem.
7. Identification of endpoints in a scalable manner.
   a) Individual — e.g., Home Smart Meter
   b) Group — e.g., All light bulbs in a room/home
   c) Scalability challenges of Individual vs. Group
   d) Sometimes the location may be more important than the individual identifier (ID)
8. Management of Multi-Party Networks
   a) For example, Smart Traffic Lights where there are several interested parties such as Emergency Services (User), Municipality (owner), Manufacturer (Vendor)
   b) Who has provisioning access?
   c) Who accepts Liability?
9. Crypto Resilience
   a) Embedded devices may outlive algorithm lifetime
   b) For example, Smart meters could last beyond 40 years
   c) Crypto algorithms have a limited lifetime before they are broken
10. Physical Protection
    a) Mobile devices can be stolen
    b) Fixed devices can be moved
11. Tamper Detection techniques and design
    a) Always On: High Poll rate, more energy, quick detection
    b) Periodic Poll: Less energy, slower detection
    c) On-event Push: Minimal energy, no detection

[Cisco, Securing the Internet of Things]

## 3. Solution for IoT Security implementation

### *3.1. Encryption Security Solution*

Encryption of information is alternate way out to protect the network from attack, which is widely used and also popular. The most common and popular algorithms used for encrypting are: RSA, ECC, AES, 3DES, MD5 and SHA, which are heavily computational [Arijit Ukil,Jaydip Sen & Sripad Koilakonda, 2011]. For each potential message, a specific code is used to check the validity of the message. By using protocols such as IPSec, the accessibility and authenticity will be provided for the data flow. For implementing these algorithms, there should be specific and dedicated processors such as Digital Signal Processors (DSP) to provide the required highly computational process. In Most cases, these processors only supply one class of encryption algorithm.

Meanwhile, in different layer, security should be provided. Which is an important aspect that should not be forgotten. In the next section, we will discuss about different ways of security requirement for IoT. [Hemant Kumar and Archana Singh, 2016]

### *3.2. Security in IoT/M2M*

As the applications of the IoT/M2M affect our daily lives, whether it is in the industrial control, transportation, Smart Grid or healthcare verticals, it becomes imperative to ensure a secure IoT/M2M system. With continuous change of IP networks, IoT/M2M applications have already become a target for attacks that will continue to grow in both magnitude and complexity. The scale and framework of the IoT/M2M make it a persuasive target for companies, organizations, nations, and more importantly people hackers. The targets of attack are abundant and cover many different industries. The possible impact could be reflected in minor irritant to significant damage to the infrastructure and loss of human life.

The device identity and mechanisms to authenticate the IoT is one of the fundamental elements in securing an IoT infrastructure. Many IOT devices may not have the required computation power, memory or storage to sustain the current authentication protocols. There are various strong encryption and authentication schemes are used today, which are based on cryptographic suites such as Advanced Encryption Suite (AES) for confidential data transport, Rivest-Shamir-Adleman (RSA) which is used in digital signatures and key transport and Diffie-Hellman (DH) for key negotiations and management. As the protocols are robust, they require high compute platform, a resource that may not exist in all IoT-attached devices. Therefore, authentication and authorization will require appropriate re-engineering to accommodate our new IoT connected world.

These authentication and authorization protocols require user-intervention in terms of provisioning and configuration. However, many IoT devices will have limited access, thus required initial configuration to be protected from theft, tampering and other forms of compromise throughout its usable life, which in many cases could be years.

These issues can be overcome with new authentication schemes which can be built using the experiences of today's strong encryption/authentication algorithms. New technologies and algorithms are being worked on. As an example, the National Institute of Standards and Technology (NIST) has chosen the compact SHA-3 as the new algorithm for the so-called "embedded" or smart devices that connect to electronic networks but are not themselves full-fledged computers [Wiki].

Some security elements that could be considered as following:

- Geographic location and privacy levels to data to be applied.
- Strong identities
- Strengthening of other network-centric methods such as the Domain Name System (DNS) with DNSSEC and the DHCP to prevent attacks
- Adoption of other protocols that are more tolerant to delay or transient connectivity (such as Delay Tolerant Networks) [Wiki]

There are many security considerations for IoT protocols depends on encryption. As new workflows emerge for sensors and components connected to the web, a disparity in time horizons creates an additional gap: devices might outlive the encryption effectiveness. For example, a power meter in a home may last fifty years, whereas the encryption protocol might survive half of that time before it is compromised. [Securing the Internet of Things: A Proposed Framework, Cisco]

The devices of IoT must use secure communication and the data transport channels to allow sending and collecting data to and from the agents and the data collection systems. While not all IoT endpoints may have bi-directional communications, leveraging message (automatically or via a network administrator) allows secure communication with the device when an action needs to be taken. [Securing the Internet of Things: A Proposed Framework, Cisco]

## 4. DNA encoded ECC for IoT security

Elliptic curve cryptography were discovered by Neal Koblitz and Victor Miller in 1985 [N. Koblitz, 1987][V.S. Miller, 1985]. ECC is the most efficient public key encryption method based on the concept of elliptic curve which is used for enhanced cryptographic key. Generally, Elliptic Curve Cryptography is used to compare with the public key encryption systems like RSA and diffie-hellman key exchange problem. ECC helps to provide highest security with low power computing devices. Some public key encryption systems like RSA, D-H key exchange and Digital Signature Algorithm (DSA) are very suitable for high power computation but when we choose IoT or cloud computing then the low power computing devices may not support such types of devices. [Hemant Kumar and Archana Singh, 2016]

**Table 1 - Key size and key size ratio between ECC and RSA/DSA.**

[Bafandehkar, Mohsen & Yasin, Sharifah & Mahmod, Ramlan & Zurina, Mohd Hanapi. (2013)]

| ECC Key Size | RSA/DSA Key Size | Key Size Ratio (ECC/RSA) |
|---|---|---|
| 106 | 512 | 1:5 |
| 132 | 768 | 1:6 |
| 163 | 1024 | 1:7 |
| 192 | 1536 | 1:8 |
| 210 | 2043 | 1:10 |
| 256 | 3024 | 1:12 |
| 384 | 7680 | 1:20 |
| 521 | 15360 | 1:30 |

### *4.1. Why use DNA encoded ECC in IoT?*

To protect private communication and keep it private from everyone except the intended recipients cryptography is the used. To securing IoT devices and many other electronic products proper use of cryptographic technique is essential. In the blog titled, "Robust IoT security costs less than you think," shown that the effectiveness of IoT security can be accomplished by a cryptographic IC that adds less than $1 to an IoT device bill of material (BOM). Implementing public-key cryptosystem for IoT devices requires significant expertise, but organizations may hire IoT security partners who are capable of providing complete, robust security solutions, typically in a matter of weeks, that is generally performed in parallel with different development efforts.

Obviously, any key will in theory be broken by a brute-force attack with sufficient computing power. The practical approach of contemporary cryptography is to use a key of adequate enough length that it can't be broken without an extraordinary amount of computing power that would be significantly more than the value of the contents that the cryptosystem protects. In the above mentioned sub-$1 cryptographic IC utilizes 256-bit Elliptical Curve Cryptography (ECC) keys, which is so secure that the computational power to break a single key may require computing resources equal in cost to 300 million times the entire world's annual GDP (78 trillion USD) working for an entire year.

.

IoT product maker has no valid excuse to continue to ignore IoT security risks because the modern cryptographic ICs make IoT security most affordable.

### *4.2. DNA Cryptography- Biological Background*

DNA is the genetic material of eukaryotes, with a double-helix molecular structure and two single-strands parallel to each other. DNA is called a polymer and composed of many small nucleotides. Each nucleotide consists of three parts:

1. The Nitrogenous bases;
2. Deoxyribose;
3. Phosphate.

DNA is the abbreviation of deoxy-ribo nucleic acid which is the germ plasma of all life. DNA is a kind of biological macro molecule and is made up of nucleotide. Each of the nucleotide contains a single base and there are four types of bases, named as Adenine(A) and Thymine(T) or Cytosine(C) and Guanine(G) corresponding to four kinds of nucleotides.

A single-stranded DNA is constructed with the following orientation: one end is called 5 and the other end is called 3. In nature DNA usually exists as double-stranded molecules. The two complementary DNA strands are held together to form a double helix structure by hydrogen bonds between the complementary bases of A and T (or C and G). [Er.Ranu Soni, Er.Vishakha Soni and Er.Sandeep Kumar Mathariya, 2011]

### *4.3. DNA Cryptography Technique*

DNA cryptography is one of the study of DNA computing about how to use DNA as an information carrier and to use mathematical operation to transfer plaintext into cipher text. Some DNA Cryptography techniques used in[H. Z. Hsu and R. C. T. Lee] will be described in this section.

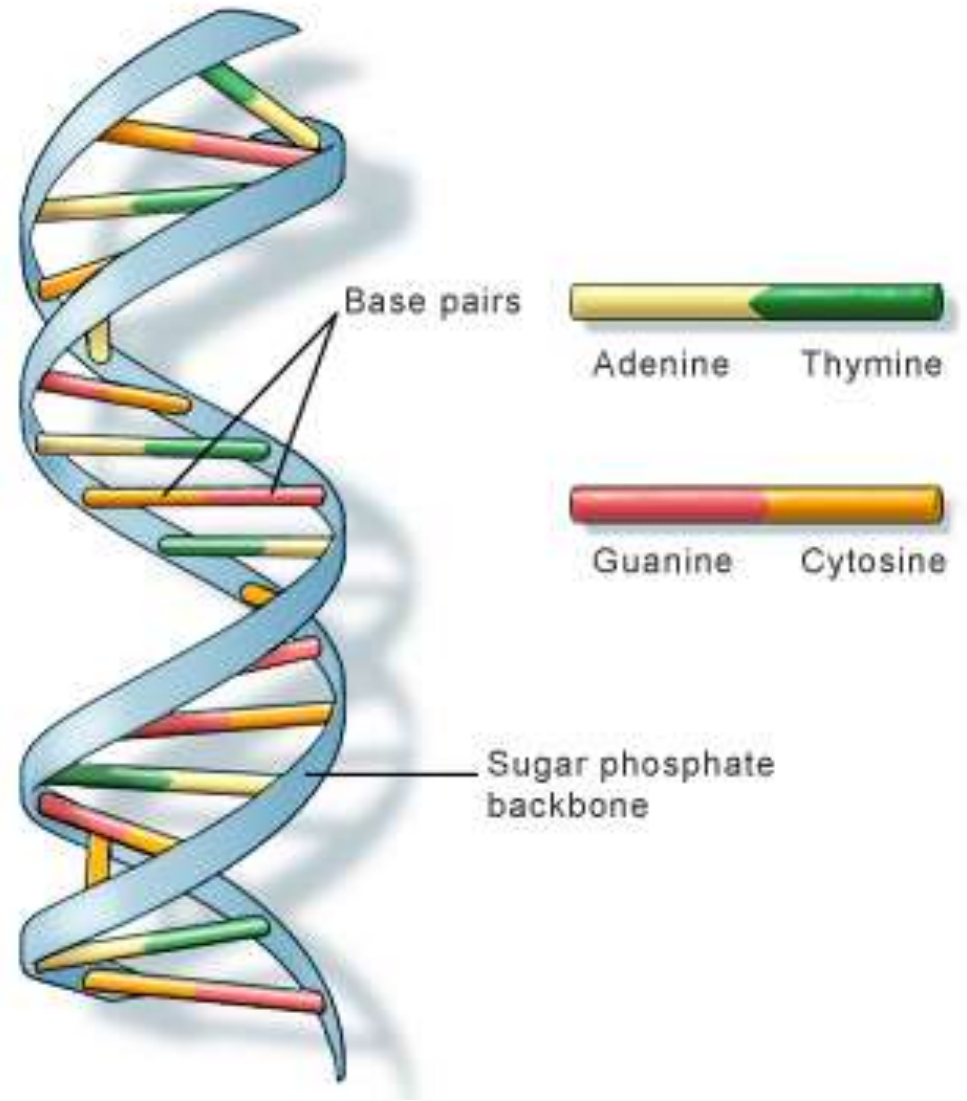

**Fig. 1 – DNA Structure.**

For encryption and decryption operation the sequence of DNA base is used in DNA cryptography. A DNA base sequence consists of four alphabets: A, C, G and T. Each alphabet is related to a nucleotide. The DNA sequence is usually quite long. We shall use part of the long DNA sequence for our example. The following are two DNA sequences are taken for our example.

The first given sequence one is a segment of DNA sequence of Litmus, where as its real length is with 2856 nucleotides long:
ATCGAATTCGCGCTGAGTCACAATTCGCGCTGA GTCACAATT
CGCGCTGAGTCACAATTGTGACTCAGCCGCGAATTCCTGCAGC
CCCGAATTCCGCATTGCAGAGATAATTGTATTTAAGTGCCTAGC
TCGATACAATAAACGCCATTTGACCATTCACCACATTGGTGTGC
ACCTCCAAGCTCGCGCACCGTACCGTCTCGAGGAATTCCTGCA
GGATATCTGGATCCACGAAGCTTCCCATGGTGACGTCACC
[European Bioinformatics Institute].

The second given sequence is a segment of DNA sequence of Balsaminaceae, where as its real length is with 2283 nucleotides long:
TTTTTATTATTT TTTTTCATTTTTTTCTCAGT TTTTAGCAC
ATATCATTACATTT TATTTTTTCATTACTTCTAT CATTCTAT
CTATAAAATCGATT ATTTTTATCACTTATTTTTC TAATTTCC
AATATTTCATCTAA TGATTATATTACATTAAAGA AATCGGTT
AAAAGCGACTAAAA ATCAATCTGGAACAAGGCTT AGTTTATT
TAATATATTATTTT ATGTAATTTCTATTGAAAAA TTAGTTAA
AAGGCAAGTATTT GAGAT [European Bioinformatics Institute].
There are a large number of DNA sequences publicly available in various web-sites, such as[European Bioinformatics Institute]. Publicly available DNA sequences are to be around 55 million[European Bioinformatics Institute]. Three DNA based encryption methods are designed in[H. Z. Hsu and R. C. T. Lee]. These methods secretly select a reference sequence S from publicly available DNA sequences. This reference sequence is only known by the sender and the receiver. This selected DNA sequence S is transformed by the sender into a new sequence S' by incorporating the DNA sequence S with the secret message M. The transformed sequence S' is sent to the receiver together with many other DNA sequences by a sender. The receiver examine all of the received sequences, recognize S' and recover the secret message M[H. Z. Hsu and R. C. T. Lee]

### *4.4. DNA Cryptography Operation*

Three DNA cryptography methods mentioned in [H. Z. Hsu and R. C. T. Lee] are generally used. The methods are as follows

1. Insertion method.
2. Substitution Method.
3. Complementary Pair approach.

In all of above the techniques, a common method of encoding and decoding is used. The plain text is converted to binary numbers. The converted binary numbers are then converted into equivalent DNA nucleotides sequence [Prokash Barman & Banani Saha, 2015]. Then one of the DNA-cryptographic methods of [H. Z. Hsu and R. C. T. Lee] is used for encryption or decryption.

The encoding and decoding operation are based on the following facts:
The four basic units of DNA, are encoded into binary in the following manner:

**Table 2 - DNA Nucleotide to Binary conversion table.**

| DNA Nucleotide Base | Binary equivalent |
|---|---|
| Adenine (A) | 00 |
| Thymine(T) | 01 |
| Guanine(G) | 10 |
| Cytosine(C) | 11 |

Plain texts are converted into binary, in the encoding phase of insertion method of DNA cryptography. A DNA sequence is chosen from publicly available sequences. The DNA sequence is converted into binary following Table -2. Then divide the binary DNA sequence into segments, where each segment contains a randomly selected number of bits. Randomly selected number of bits should be greater than 2. At the beginning of segmented binary DNA sequence, each bit of binary plain text is then inserted. Then the inserted bit sequences are concatenated to obtain new encoded binary sequence. A new pseudo (not found really, only obtained from operation) binary sequence is obtained by converting the encoded binary sequence into nucleotide using Table -2. [Prokash Barman & Banani Saha, 2015]

### *4.5. DNA Encoded Elliptic Curve Cryptography Scheme*

***ECC-DNA Cryptosystem***

The elliptic curve cryptography and DNA cryptography are the most modern cryptographic technique. Thus, for better security, characteristics of both the systems may be combined to construct highly secured cryptographic techniques. In the new system for encoding of plain text we propose to use the insertion method of DNA cryptography as used in[H. Z. Hsu and R. C. T. Lee]. The plain text is converted to its equivalent ASCII value. Then the ASCII values are converted into binary. From publicly available sequences of DNA, we choose known DNA nucleotide sequence. Both the sender and the receiver should synchronize with the chosen DNA sequence. The DNA nucleotide sequences are converted into binary using table – 2. In this stage we will obtain several pairs of binary numbers. All the binary number pairs are concatenated to make a long binary number. Then the obtained binary number is broken into several segments. Here an arbitrary number of bits, greater than 2 (it was already pairs of bits in early stage) will be taken in each segment. Now each bit of converted binary plaintext inserted into the beginning of the binary segments of nucleotide sequence. The segments are concatenated again and converted to Nucleotide letter. (A,T,G & C). Now the new sequences are converted into decimal following conversion rules in table – 2. In our proposed system this steps are used for encoding. For encryption the decimal numbers are converted into elliptic curve point using Koblitz method[H. Z. Hsu and R. C. T. Lee]. This point is called plain test point. The points are encrypted into another elliptic curve point using ECC encryption expression (1). ECC encryption process is done with the help of its generated keys. The encrypted points on the elliptic curve are called cipher text points. This point is send to the receiver. In the reverse process the receiver will derive the plain text message.

$\{kG, P_m + k\,P_B\}$ (Expression-1)
Where,
G - Generated Points
$P_m$ - Plaintext points
k - Random number chosen by user

PB - Public key of another user
Pm + kPB - nB (kG) = Pm + k (nB) G - nB (kG) = Pm (Expression-2)

The cipher text points are deciphered using ECC decryption expression (2). Deciphered points are converted into numbers using Koblitz's method. These numbers are decoded to an unknown DNA nucleotide sequence. From the unknown DNA sequence, known DNA nucleotide sequence (S) is decoded to obtain plaintext.

**Table 3 - Nucleotide to number conversion table.**

| DNA Nucleotide Base | Binary equivalent |
|---|---|
| A | 10 |
| T | 20 |
| G | 30 |
| C | 40 |

### *4.6. DNA-ECC Cryptographic Algorithm*

Input: Plain text (P), number of bits of DNA sequence segment (k), known DNA sequence (D)
Output: Cipher text points(C)}
Procedure
Begin
Input P
Convert P into Binary P'
Convert D into Binary D'
Segment D' with k bit in a segment
Insert each bits of P' into the beginning of each segment of D'
Concatenate segments of D'
Convert D' into DNA nucleotide DN (where A=00, T=01, G=10, C=11)
Convert DN into decimal N (follow Table-3)
Call koblitz to convert N into ECC point and cipher text C
End
end Procedure

[Prokash Barman & Banani Saha, 2015]

### *4.7. Example for converting plain text*

Plain text message (P): "t"
ASCII message: 116
Binary message(P'): 1110100

DNA sequence (D):
ATCGAATTCGCGCTGAGTCACAATTCGCGCTGAGTCACAATTC
GCGCTGAGTCACAATTGTGACTCAGCCGCGAATTCCTGCAGCC
CCGAATTCCGCATTGCAGAGATAATTGTATTTAAGTGCCTAGCT
CGATACAATAAACGCCATTTGACCATTCACCACATTGGTGTGC
ACCTCCAAGCTCGCGCACCGTACCGTCTCGAGGAATTCCTGCA
GGATATCTGGATCCACG AAGCTTCCCATGGTGACGTCACC

- Binary DNA sequence (D'): 00 01 11 10 00 00 01 01 11 10 11 10 11 01
- Segmented Binary DNA sequence (where k=3): 000 111 100 000 010 111 101 110 110 1
- Insert each bits of P' into beginning of each segments of D': 1-000 / 1-111 / 1-100 /0-000 / 1-010 / 0-111/ 0-101 / -110 /-1
- Concatenate the segments of D': 100011111100000010100111101011101
- Convert D' to nucleotide DN
  10-00- 11-11-11- 00- 00- 00-10-10- 01-11-01- 01-11-01
  G- A- C- C- C- A- A- A- G- G- T- C- T- T- C- T
- Convert DN to ASCII
  G A C C C A A A G G T C T T C T
  30 10 40 40 40 10 10 10 30 30 20 40 20 20 40 20
- Covert ASCII(DN) to Elliptic Curve point which is encrypted cipher text point

## 5. Need of resilient mechanism in IoT

In the context of smart cities the ICTs have the important role to provide the infrastructure to deploy the devices and services, thus it is vital to keep this infrastructure alive. With the purpose to improve the robustness and reliability of the services in smart cities, it is needed a resilient IoT framework capable to react and self-configure after unpredictable events. This important requirement, include resilience in IoT infrastructure, is one of the open issues in the IoT field that needs to be tackled. [Rodrigo Roman, Pablo Najera and Javier Lopez, 2011]

These issues have particular significance in the IoT where the availability of secure data is most important. For example, an industrial process may rely on timely and accurate temperature measurement for its operations. If that industrial process endpoint is undergoing a Denial of Service (DoS) attack, the process collection agent for temperature measurement, must somehow be made aware about the DoS. In such event, the IoT system should be able take appropriate alternative actions in real-time, like as sourcing data from a secondary connection, or delay the information transmission. The system must also be able to distinguish between loss-of-data due to an on-going DoS attack and loss of data due to a catastrophic device disruption event in the plant. It might accomplish this by using learning machine techniques (for example, comparing a normal operational state to an attack state previously learned) [Cisco, Securing the Internet of Things].

## 6. Conclusion

In this paper we describe a novel cryptographic scheme by combining DNA Encoding with ECC algorithm. DNA Encoding with Elliptic curve cryptography has various advantages over traditional systems, in the form of increased speed, less memory and smaller key size. In addition, the proposed system has two level security - one in DNA encoding and other in ECC encryption steps. On the principle described in this paper a DNA-ECC embedded system may be developed using FPGA based embedded system to verify the practical implementation of the proposed cryptographic scheme.